\documentclass[11pt]{article}

\usepackage[verbose=true,letterpaper]{geometry}
\AtBeginDocument{
  \newgeometry{
    textheight=9in,
    textwidth=6.5in,
    top=1in,
    headheight=14pt,
    headsep=25pt,
    footskip=30pt
  }
}
\usepackage{setspace} 
\usepackage{graphicx}
\usepackage{amsmath}
\usepackage{cite}
\usepackage{hyperref}       
\usepackage{url}            
\usepackage{booktabs}       
\usepackage{amsfonts}       
\usepackage{nicefrac}       
\usepackage{microtype}      
\usepackage{lipsum}
\usepackage[hang,flushmargin]{footmisc}
\usepackage{multibib}
\newcites{supp}{References}

\begin{document}

\parindent 0em

{\Large\textbf{Melting curves of ice polymorphs in the vicinity of the liquid-liquid critical point}}

\vspace{1cm}

{ \large
Pablo M. Piaggi$^{1,\dagger}$, Thomas E. Gartner III$^{2,\dagger}$, Roberto Car$^{1,3}$, and Pablo G. Debenedetti$^{4,*}$\let\thefootnote\relax\footnote{$^1$Department of Chemistry, Princeton University, Princeton, NJ 08544, USA, $^2$School of Chemical and Biomolecular Engineering, Georgia Institute of Technology, Atlanta, Georgia 30318, USA, $^3$Department of Physics, Princeton University, Princeton, NJ 08544, USA, $^4$ Department of Chemical and Biological Engineering, Princeton University, Princeton, NJ 08544, USA, $^\dagger$These authors contributed equally}\footnote{$^*$e-mail: pdebene@princeton.edu} \\
}


\section*{Abstract}
The possible existence of a liquid-liquid critical point in deeply supercooled water has been a subject of debate in part due to the challenges associated with providing definitive experimental evidence.
Pioneering work by Mishima and Stanley [Nature 392, 164 (1998) and Phys.~Rev.~Lett. 85, 334 (2000)] sought to shed light on this problem by studying the melting curves of different ice polymorphs and their metastable continuation in the vicinity of the expected location of the liquid-liquid transition and its associated critical point.
Based on the continuous or discontinuous changes in slope of the melting curves, Mishima suggested that the liquid-liquid critical point lies between the melting curves of ice III and ice V.
Here, we explore this conjecture using molecular dynamics simulations with a purely-predictive machine learning model based on \textit{ab initio} quantum-mechanical calculations.
We study the melting curves of ices III, IV, V, VI, and XIII using this model and find that the melting lines of all the studied ice polymorphs are supercritical and do not intersect the liquid-liquid transition locus.
We also find a pronounced, yet continuous, change in slope of the melting lines upon crossing of the locus of maximum compressibility of the liquid.
Finally, we analyze critically the literature in light of our findings, and conclude that the scenario in which melting curves are supercritical is favored by the most recent computational and experimental evidence.
Thus, although the preponderance of experimental and computational evidence is consistent with the existence of a second critical point in water, the behavior of the melting lines of ice polymorphs does not provide strong evidence in support of this viewpoint, according to our calculations.



\newpage
\parindent 1em

\section*{Introduction}
\label{sec:introduction}
Water continues to be the focus of intense scientific inquiry, not only because of its importance in the biological and physical sciences, but also on account of its distinctive thermophysical properties and phase behavior. Water exhibits at least 17 different crystalline phases (with new ones continuing to be uncovered) \cite{Salzmann19,Hansen21}, multiple glassy states \cite{Loerting11}, and possibly also a liquid-liquid phase transition (LLT) between high-density and low-density liquids (HDL and LDL, respectively) under supercooled conditions \cite{Poole92,Gallo16}. As such, water provides a rich proving ground to stretch our understanding of diverse thermophysical phenomena including complex phase equilibria, metastable phase transitions, and glass physics \cite{DebenedettiBook}, as well as the possible relationships between them \cite{Handle17,Debenedetti03}. 

The possibility of an LLT in water has been the focus of numerous studies \cite{Gallo16}, and a preponderance of both experimental and computational evidence points to the existence of water's LLT at positive pressures ($P$) and supercooled temperatures ($T$) (i.e., below the melting $T$ of the stable ice I phase) \cite{Kim20,NILSSON22,Palmer14,Debenedetti20,Palmer18,Gartner22,weis2022liquid}. However, there remain many unresolved questions around the LLT and its relationship to water's properties and various solid phases. A set of observations instrumental to the development of the argument in favor of the LLT came about when Mishima and Stanley characterized the melting of various ice polymorphs to liquid water upon decompression at different $T$ \cite{Mishima98,Mishima00}. They observed that the melting curve of ice III exhibited a notable but continuous change in slope in the $T-P$ plane, while ice V and ice IV exhibited sharp and seemingly discontinuous changes in slope. Recall that, by the Clausius-Clapeyron equation \cite{callen1998thermodynamics},
\begin{equation}
\frac{dP}{dT} = \frac{\Delta H}{T_m \Delta V},
\label{eq:Clausius-Clapeyron}
\end{equation}
the slope $dP /dT$ of a line of phase coexistence $T_m(P)$ is related to the change in enthalpy $\Delta H$ and volume $\Delta V$ across the transition. This idea suggests that if a melting curve exhibits a discontinuous change in slope, it correspondingly reflects a discontinuous change in the properties of ice and/or liquid water at that point. Given that the enthalpy and volume of crystalline solids is only weakly dependent on $T$ and $P$, Mishima and Stanley concluded that the properties of the liquid phase were changing discontinuously (i.e., evidence of an LLT). This argument, if correct, would place the liquid-liquid critical point (LLCP) somewhere in between the ice V and ice III melting lines, with the LLT coexistence line intersecting the ice V and ice IV melting curves at the point of discontinuous change in slope. Mishima also probed the melting lines of ices VI and XIII, but was unable to extend those curves far enough to intersect with the possible LLT line.

This rationalization for the observed trends, while plausible, remains difficult to definitively explore experimentally due to rapid crystallization of the stable ice I phase upon melting of the other polymorphs. Similar practical challenges also hamper direct experimental demonstration of the LLT. Thus, open questions remain about the true relationship between a possible LLT and the metastable melting of the ice phases. Molecular modeling represents an attractive route to probe these ideas, as one can design simulation methodologies free from unwanted crystallization, which allow us to directly study the relationship between the LLT and the various ices. In parallel, advances in machine learning (ML)-based interaction potentials \cite{Noe20,Wen22} allow us to develop predictive intermolecular potential models that describe water's interactions at the level of an \textit{ab initio} reference calculation (e.g., density functional theory), thus enabling purely-predictive simulations of complex collective properties and phase behavior at tractable computational cost \cite{Gartner20,reinhardt2021quantum,Zhang21,Piaggi21,Schran21,piaggi2022homog}. In this study, we coupled one such ML-potential method (Deep Potential Molecular Dynamics, DPMD) \cite{Zhang18,Wang18} with several advanced simulation techniques to shed further light on the possible relationship between the LLT and water's liquid-solid phase behavior. 

\section*{Potential scenarios}
\label{sec:potential_scenarios}

\begin{figure*}
\includegraphics[width=\textwidth]{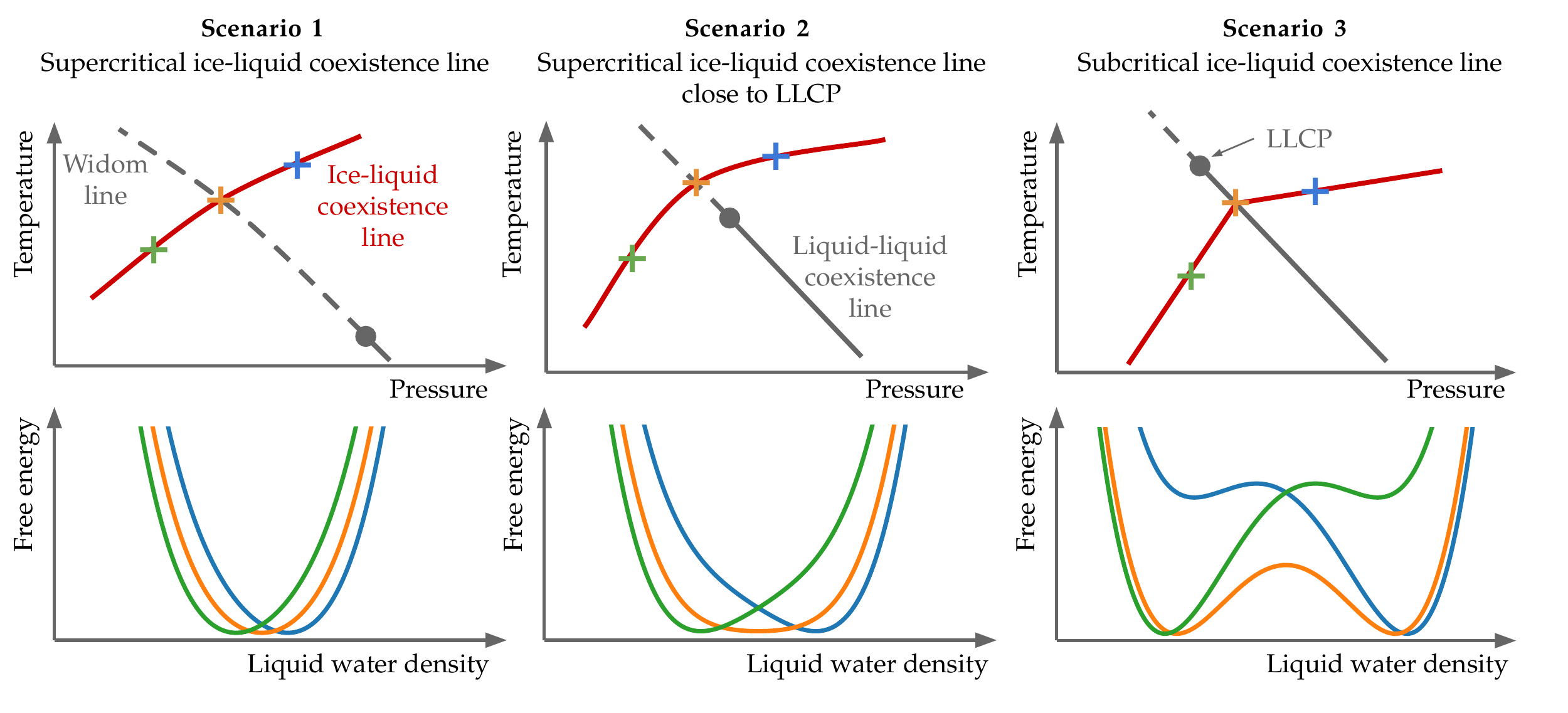}
\caption{\label{fig:Fig1} Hypothetical scenarios describing the possible relationship between ice polymorph melting curves and the LLT. The upper plots show the melting curve of a hypothetical ice polymorph (red solid line), the LLT line (gray solid line), the LLCP (gray circle), and the Widom line (gray dashed line). The lower plots show hypothetical free energy surfaces for the liquid density along the melting curves at the three points marked by \textbf{+} signs. Scenario 1 (left) shows a case where the melting curve is significantly supercritical, Scenario 2 (center) shows a case where the melting curve is slightly supercritical, and Scenario 3 (right) shows a case where the melting curve is subcritical.}
\end{figure*}

Before describing the details of our approach and results, we illustrate schematically the possible classes of behavior in FIG.~\ref{fig:Fig1}.
In this discussion, we assume the existence of an LLT.
The elements that we consider in our analysis are the melting curve of an ice polymorph, the liquid-liquid critical point, the liquid-liquid coexistence line (or binodal), and the Widom line. The Widom line can be regarded as an extension of the liquid-liquid coexistence line to supercritical conditions and is defined by the locus of maxima of the correlation length.
Response functions, such as the heat capacity at constant pressure $C_P$ and the isothermal compressibility $\kappa_T$, also have pronounced maxima at supercritical conditions even far from the critical point, and the values of the response functions diverge as the critical point is approached \cite{xu2005relation}.
Furthermore, the lines of maxima of the response functions in the $T-P$ plane asymptotically converge to the Widom line as the critical point is approached from supercritical conditions \cite{xu2005relation}.
$C_P=(\partial H/ \partial T)_P$ and $\kappa_T=-(1/V)(\partial V/\partial P)_T$ are derivatives of the enthalpy $H$ and volume $V$, and thus we expect the fastest change in these liquid-state properties in the immediate vicinity the Widom line.
In turn, a pronounced change in the enthalpy and volume of the liquid at the Widom line will lead to correspondingly pronounced changes in slope of the ice melting line as predicted by Eq.~\eqref{eq:Clausius-Clapeyron}.

We now analyze three possible scenarios.
If the melting curve of a particular polymorph were to be significantly supercritical (Scenario 1, FIG.~\ref{fig:Fig1} left), the impact of the critical point would be minimal.
Therefore, we would expect to observe a modest change in slope of the melting curve and the free energy surface of the liquid state would have a single basin that smoothly moves from high to low density as temperature decreases along the melting curve. If the melting curve passed near to the critical point but still at supercritical conditions (Scenario 2, FIG.~\ref{fig:Fig1} center), a more significant but still continuous change in slope might be observed as the liquid properties change swiftly but continuously upon crossing the Widom line. In this case, the free energy surfaces would still only show one single minimum at a given state point yet they can show significant asymmetry \cite{Gartner22}, and broadening at the intersection of the melting curve with the Widom line.
The broadening of the free energy surface of the liquid as a function of density at the Widom line follows from the fact that density fluctuations $\sigma_{\rho}$ are related to $\kappa_T$ via $\sigma_{\rho}^2=\rho^2 k_B T \kappa_T/ V$ where $\rho$ is the density and $k_B$ the Boltzmann constant \cite{pathria2016statistical}.
Finally, if the melting curve was subcritical (Scenario 3, FIG.~\ref{fig:Fig1} right), a discontinuous change in liquid properties across the LLT would result in a discontinuous change in the slope of the melting curve, and a free energy surface with two basins of equal depth would develop at the point of liquid-liquid phase coexistence (i.e., where the ice melting line meets the LLT line). Moving forward, we will situate our simulation results in the context of these three potential scenarios.

\section*{Calculation of melting curves}
\label{sec:methods1}

Our molecular dynamics simulations were driven by a deep potential model\cite{Zhang18} of water developed by Zhang et al. \cite{Zhang21}
The model has been carefully trained to reproduce with high fidelity the potential energy surface of water based on density functional theory (DFT) calculations with the Strongly Constrained and Appropriately Normed (SCAN) exchange and correlation functional \cite{Sun15}.
SCAN is one of the best semilocal functionals available and describes with good accuracy many properties of water and ice, and their anomalies \cite{Sun16,Chen17,Piaggi21}.
Even though the model is short-ranged with a cutoff of 6 \AA, it can capture subtle physical effects, such as polarization \cite{piaggi2022homog} and many-body correlations \cite{Zhang18}.
Furthermore, this model describes qualitatively the behavior of water and ice polymorphs in a region of the phase diagram spanning temperatures 0-500 K and pressures 0-50 GPa \cite{Zhang21}.
It is thus suitable to represent ice III, IV, V, VI, and XIII at the conditions of interest for this work.
Another aspect of critical importance is whether the model has a liquid-liquid transition at deeply supercooled conditions.
We recently proved rigorously using free energy calculations that this model has a liquid-liquid transition with a critical point at $T_c = 242 \pm 5$ K and $P_c = 0.295 \pm 0.015$ GPa \cite{Gartner22}.
It is important to note that SCAN also has limitations.
Largely due to the self-interaction error in semilocal functionals \cite{sharkas2020self}, the strength of the hydrogen bond is overestimated, resulting in an upward displacement of melting temperatures of about 40 K with respect to experiments \cite{Piaggi21}. Additionally, the solid polymorphs ice III and ice XV are incorrectly predicted by SCAN to be metastable at all ($T$, $P$) \cite{Zhang21}. However, given the complexity of water's phase diagram, SCAN predicts the relative location of the various phase boundaries in good agreement with experiment \cite{Zhang21}.

\begin{figure*}
\includegraphics[width=0.95\textwidth]{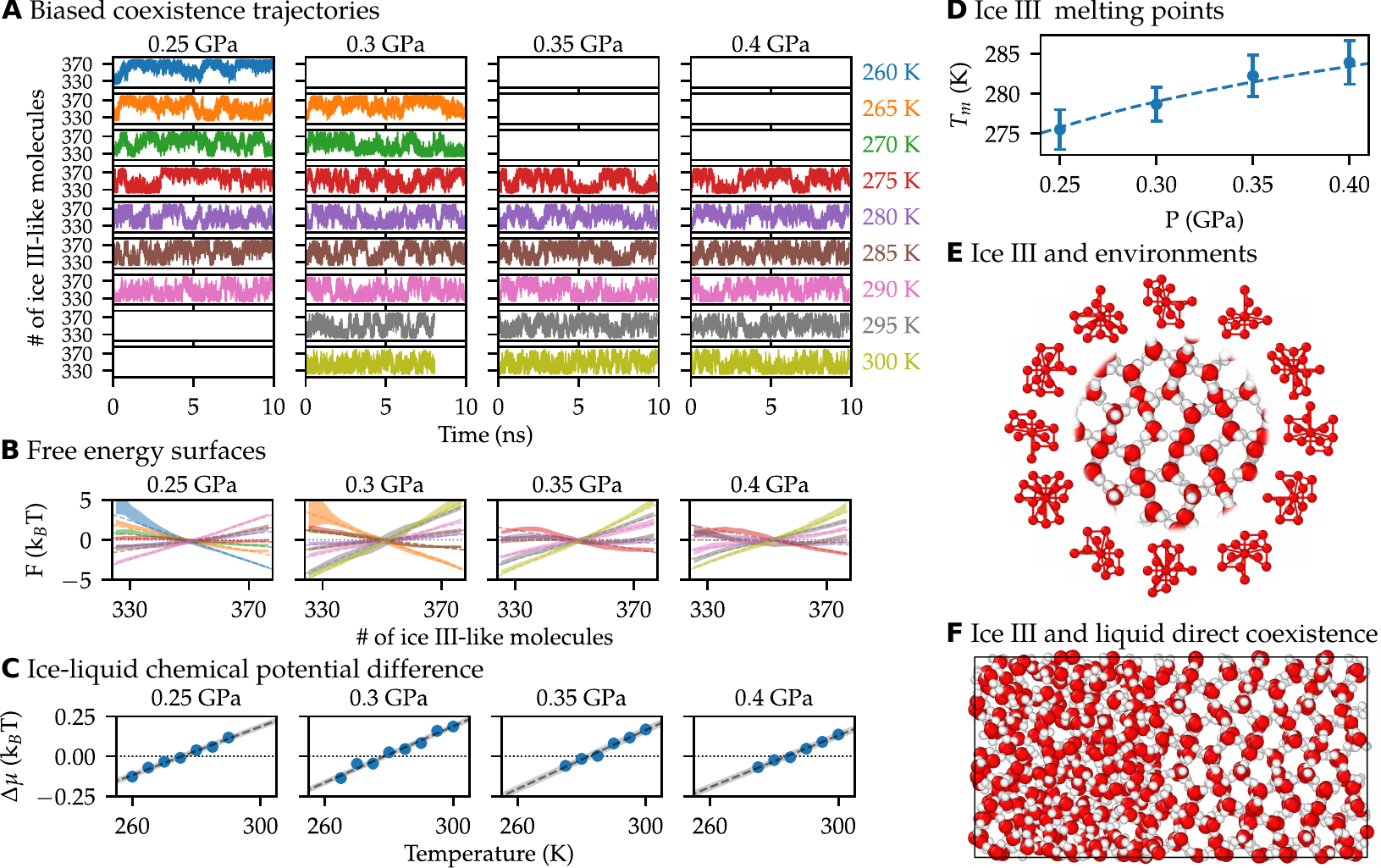}
\caption{\label{fig:Fig2} Overview of the methodology to calculate melting curves of ice polymorphs. The procedure is illustrated using the case of ice III. (A) Number of ice III-like molecules as a function of time in the biased coexistence simulations at various $T$ and $P$. The colors of the curves correspond to the $T$, as labeled to the right of the figure. Empty plots denote that no simulations were run at that ($T$, $P$). The range (324,378) that is reversibly sampled corresponds to one layer of ice III. (B) Free energy surfaces as a function of number of ice III-like molecules, where the dashed line is a linear fit to the free energy surface and the shaded region denotes the uncertainty. Colors match the same $T$ reported in panel (A) above. (C) Chemical potential difference between ice III and liquid at various $T$ and $P$. The gray dashed line is a linear fit to the data, and the shaded region represents one standard deviation of uncertainty in the fit parameters. (D) Melting curve obtained by this procedure, where the blue points represent the $T$ and $P$ of zero chemical potential difference between ice and liquid obtained in panel (C). Error bars represent one standard deviation errors in the fit parameters as shown in (C). The dashed line is the melting curve obtained from the integration of the Clausius-Clapeyron equation. (E,F) Simulation snapshots illustrating ice III and the molecular environments used to generate the order parameter \cite{Piaggi19,Bore22} to drive the biased coexistence (E), and the ice III-liquid coexistence geometry (F).}
\end{figure*}

Herein, we computed the melting lines of the ice polymorphs in two stages.
In the first stage, we calculated a few points along the liquid-solid coexistence lines using a biased coexistence approach \cite{Bore22} in which we simulate a particular ice polymorph and liquid water in direct coexistence (FIG.~\ref{fig:Fig2}F), and use a bias potential to reversibly crystallize and melt a layer of solid (FIG.~\ref{fig:Fig2}A).
This approach was used in a recent work to calculate the phase diagram of the state-of-the-art empirical model of water TIP4P/Ice \cite{Abascal05}, and can be regarded as a generalization of the interface pinning approach \cite{Pedersen13}.
From biased coexistence simulations carried out at different temperatures and pressures, we extract the difference in chemical potential between the liquid and ice from the slope of the free energy surfaces\cite{Pedersen13,Bore22} (FIG.~\ref{fig:Fig2}B), and locate the liquid-ice coexistence temperature at a given pressure as the temperature at which this difference is zero (FIG.~\ref{fig:Fig2}C-D).
We applied this procedure to ice III, IV, V, and XIII to obtain a few coexistence points for each polymorph.
See FIG.~\ref{fig:Fig2} for an overview of this procedure for the case of ice III.
We show the results for ice IV, V, and XIII in the Supplementary Material \cite{SI}.
We also validated the coexistence points obtained via the biased coexistence method for ice IV and V using standard direct-coexistence simulations (see the Supplementary Material \cite{SI}).
We subsequently obtained continuous and smooth coexistence lines by integrating the Clausius-Clapeyron equation as first proposed by Kofke \cite{Kofke93}.
This technique is based on the numerical integration of Eq.~\eqref{eq:Clausius-Clapeyron} using the enthalpy and volume obtained from constant temperature and pressure simulations of each phase (see Methods section and the Supplementary Material\cite{SI} for further details).

\section*{Results}

\begin{figure*}
\includegraphics[width=\textwidth]{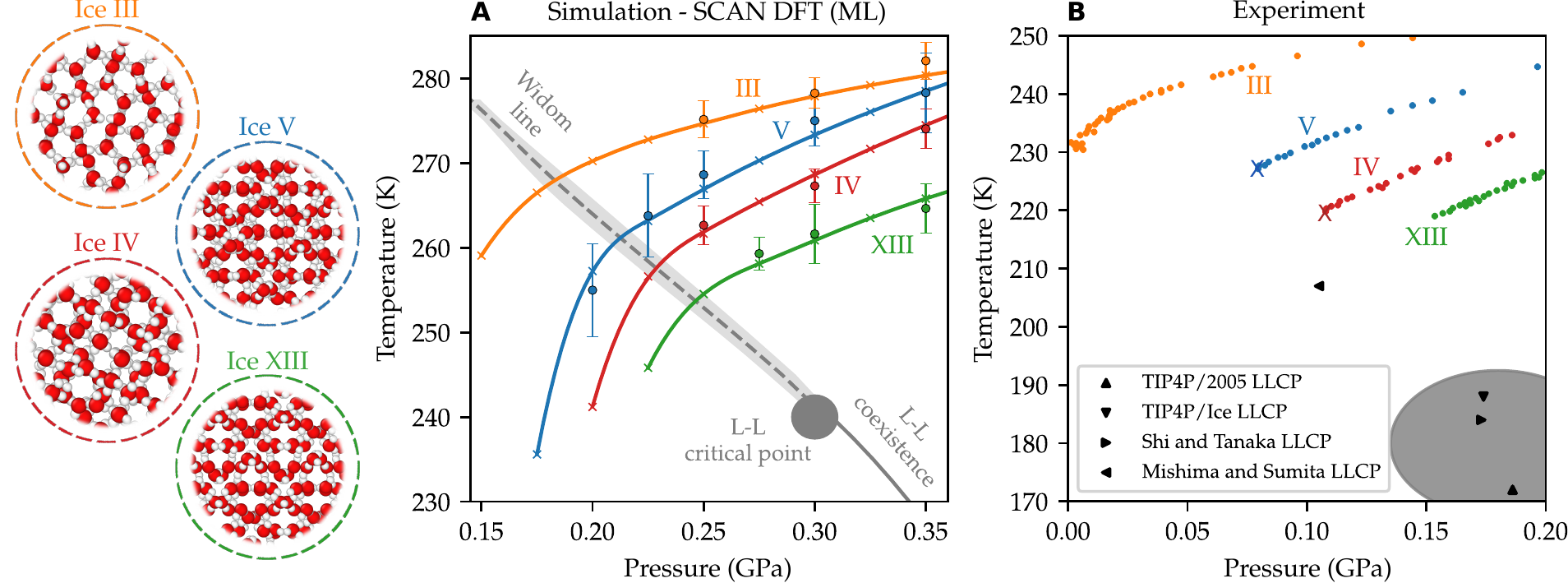}
\caption{\label{fig:Fig3} Melting curves of ice polymorphs III, IV, V, and XIII, and their location relative to the liquid-liquid critical point. A) Results obtained using a machine learning model based on the SCAN DFT functional. Circles represent melting points calculated using biased coexistence simulations \cite{Bore22}, crosses were obtained by integrating the Clausius-Clapeyron equation, and lines are spline interpolations of the latter results. We also show the location of the critical point, the liquid-liquid coexistence line, and the Widom line (line of maxima of $\kappa_T$) as calculated in our previous work \cite{Gartner22}. B) Melting curves reported by Mishima \cite{Mishima00} for heavy water based on decompression-induced melting experiments. The approximate location of the discontinuous change in slope in the melting curves of ice IV and V is marked with an X. The shaded region is the location of the critical point estimated by Bachler et al.~\cite{Bachler21}. We also show the location of the critical point obtained by Shi and Tanaka using experimental measurements \cite{Shi20}, by Debenedetti et al.\ using molecular simulations with the empirical water models TIP4P/2005 and TIP4P/Ice \cite{Debenedetti20}, and by Mishima and Sumita\cite{mishima2023equation} using an extrapolation based on polynomial fits to equation of state data. On the left, we show atomic configurations representative of ices III, IV, V, and XIII.}
\end{figure*}

Using the techniques described above, we calculated the coexistence points and lines shown in FIG.~\ref{fig:Fig3}A for ice III, IV, V, and XIII.
The circles and error bars correspond to biased coexistence simulations, and the lines were computed by integrating the Clausius-Clapeyron equation.
We also show in FIG.~\ref{fig:Fig3}A the data for the liquid-liquid critical point, liquid-liquid coexistence line, and Widom line reported recently by us \cite{Gartner22}.
According to these calculations, the melting curves of all ice polymorphs as predicted by the SCAN functional are supercritical, \textit{i.e.}, they pass above the liquid-liquid critical point.
The melting line of ice VI is also supercritical and is shown in the Supplementary Material \cite{SI}.
Thus, all of them intersect the Widom line rather than the LLT line.
Our simulations result in melting curves that show a pronounced, yet continuous, change of slope upon crossing the Widom line.
This behavior is compatible with the expected change of properties of liquid water from HDL-like to LDL-like as the Widom line is traversed from high to low pressures.
Moreover, the change in slope is smoother for ice III than for the other polymorphs, consistent with an increasingly abrupt change in the properties of the liquid closer to the critical point.
The smoother change in slope of the melting curve of ice III resembles the behavior hypothesized in Scenario 1 described in FIG.~\ref{fig:Fig1} while the more abrupt change shown by ice V, IV, and XIII is reminiscent of Scenario 2.  

Our results also show good agreement between the biased coexistence simulations and the integration of the Clausius-Clapeyron equation in the HDL-like region.
On the other hand, it was not possible to perform biased coexistence simulations in the LDL-like region due to the long relaxation times of the LDL-like liquid at those thermodynamic conditions.
Indeed, even for the comparatively less expensive bulk liquid simulations for the Clausius-Clapeyron integration procedure, we needed long simulations (100 ns) of the bulk liquid in the LDL-like region for robust statistical certainty.

\begin{figure*}
\includegraphics[width=0.9\textwidth]{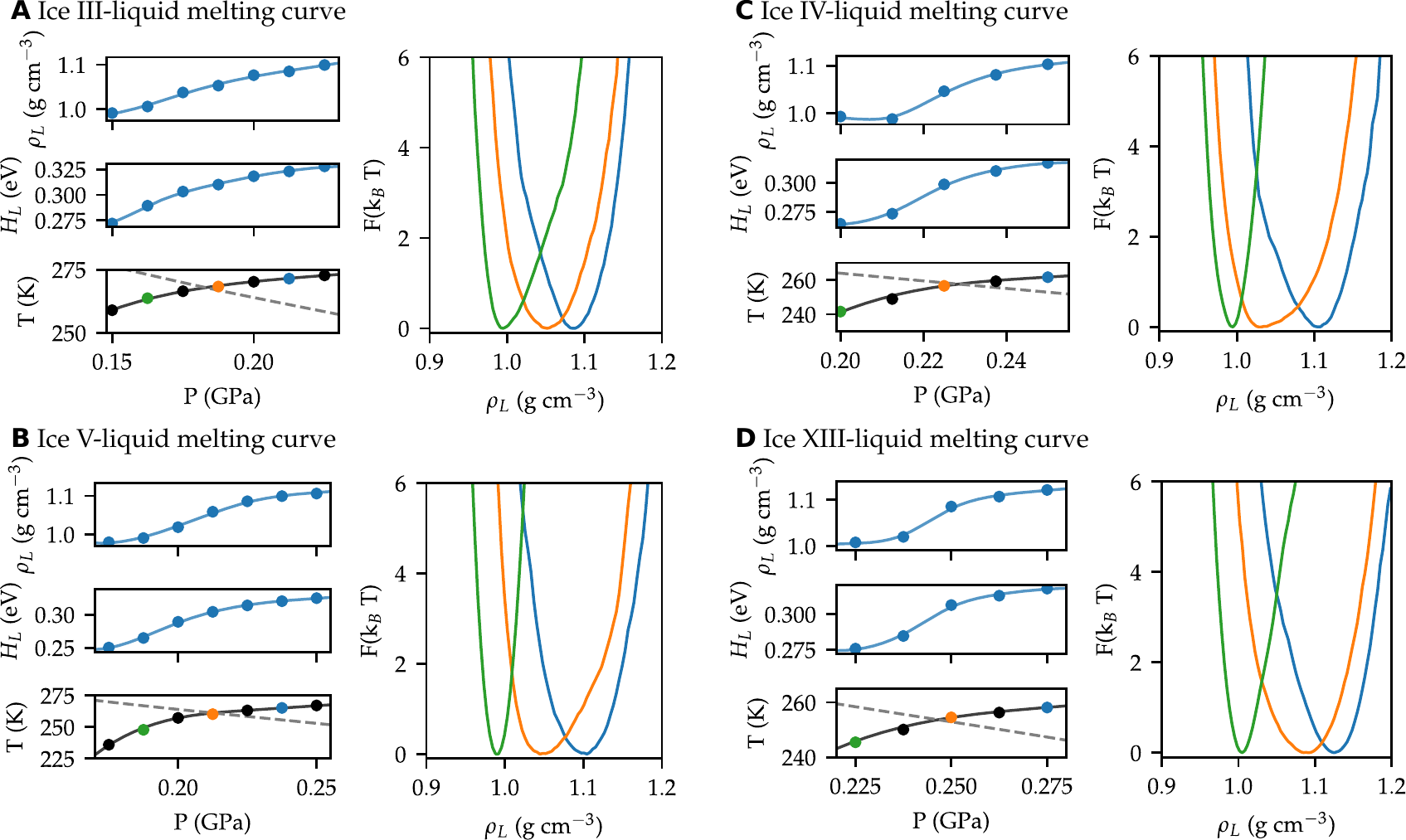}
\caption{\label{fig:Fig4} Properties of liquid water along melting curves of several ice polymorphs. Panels A, B, C, and D correspond to ice III, V, IV, and XIII, respectively. For each ice polymorph, we show the enthalpy of liquid water $H_L$, the density of liquid water $\rho_L$, and the melting temperature $T$ as a function of pressure $P$. The locus of maxima of isothermal compressibility \cite{Gartner22} is shown in the $T-P$ pane with a dashed line. We also show the free energy surfaces $F$ as a function of the density of liquid water $\rho_L$. The free energy surfaces are color-coded to match the color of points along the $T$ vs $P$ coexistence line to specify the thermodynamic conditions at which they were calculated.}
\end{figure*}

The analysis of the melting curves shown in FIG.~\ref{fig:Fig3}A does not constitute proof of a continuous change in slope since the curves are obtained from a set of points interpolated with a spline, which is by construction smooth and differentiable.
In order to provide evidence for the continuous change in slope, we now analyze in detail the properties of liquid water along the melting curves of ice polymorphs.
In FIG.~\ref{fig:Fig4} we show the enthalpy and density of liquid water as a function of pressure.
Both properties exhibit a swift change upon crossing of the Widom line and the change is more abrupt as the melting curves approach the critical point, with sequence ice III $\rightarrow$ V $\rightarrow$ IV $\rightarrow$ XIII.
We ruled out that this behavior is a result of ice crystallization by analyzing configurations at regular intervals of 5 ps.
We calculated the structural fingerprints CHILL+ \cite{nguyen2015identification} and Identify Diamond Structure \cite{Larsen16}, as implemented in Ovito\cite{Stukowski09}, and we did not find atomic environments compatible with ice I in any of our simulations.
We also show in FIG.~\ref{fig:Fig4} the free energy surfaces (FES) as a function of the liquid water density for selected points along the coexistence lines.
The FES of the liquid along the melting curves of all studied ice polymorphs show a behavior reminiscent of Scenario 2 of FIG.~\ref{fig:Fig1}.
For all ices, the FES at the state point closest to the Widom line shows clear broadening.
Furthermore, the FES in the vicinity of the Widom line exhibits deviations from a quadratic form with significant asymmetry and a shoulder suggestive of the metastable free energy minimum that would appear below the critical point.
Taken together, this behavior provides strong evidence of a continuous crossover from HDL-like to LDL-like liquids as the melting curves of ice III, V, IV, and XIII are traversed towards lower pressures.
We remark that none of the melting lines analyzed here have properties of the liquid compatible with subcritical Scenario 3 of FIG.~\ref{fig:Fig1} that would lead to a discontinuous change in the slope of the melting line.
Based on the analysis of the liquid properties described above, we conclude that the changes in slope of the melting curves shown in FIG.~\ref{fig:Fig3} are indeed continuous.

We have so far focused on the properties of the liquid phase.
However, according to Eq.~\eqref{eq:Clausius-Clapeyron}, the properties of ice can also affect the slope of melting curves.
In the Supplementary Material \cite{SI}, we show the change in enthalpy and density of ice polymorphs along the melting lines.
The data show that the changes experienced by the bulk ice polymorphs are much more subtle than the corresponding changes in the properties of the liquid phase.
In the pressure range shown in FIG.~\ref{fig:Fig4}, the densities of ice polymorphs change by less than 1\% while the density of liquid water changes by 10\%.
Furthermore, the enthalpy of ices varies by around 8\% while the enthalpy of liquid water has a significantly larger variation of around 17\%.
This analysis indicates that the changes of the properties of the liquid phase are the main factor driving the sharp changes in slope observed in FIG.~\ref{fig:Fig3}.

The results described above correspond to a purely-predictive model derived from first principles calculations.
An alternative approach is to evaluate the melting lines of ice polymorphs using semi-empirical water models that are fit to experimental information.
For this reason, we calculated the melting line of ice V in the TIP4P/Ice model \cite{Abascal05}, which is a state-of-the-art semi-empirical model for the study of ice polymorphs.
The location of the liquid-liquid critical point for this model has been determined accurately by Debenedetti et al.~\cite{Debenedetti20}.
We find that the melting curve of ice V within the TIP4P/Ice model (shown in the Supplementary Material \cite{SI}) is also supercritical, in agreement with the SCAN calculations reported above.

\section*{Discussion}

The picture that emerges from our present results is in contrast with Mishima and Stanley's interpretation \cite{Mishima98,Mishima00}.
As described above, Mishima's interpretation of the experiments considers that the melting curve of ice III is supercritical, and the melting lines of ice IV, V, and XIII are subcritical \cite{Mishima00}.
On the other hand, our calculations based on an \textit{ab initio} model predict supercritical behavior for all the studied ice polymorphs.
To evaluate this discrepancy, we analyze the consistency of each of these two interpretations in the light of the most recent evidence for the location of the critical point.
The decompression-induced melting curves measured by Mishima \cite{Mishima00} are shown in FIG.~\ref{fig:Fig3}B together with recent estimates of the location of the liquid-liquid critical point.
The estimates include an extrapolation by Bachler et al. based on experimental data for the high- and low-density spinodals obtained from compression/decompression experiments on glassy water \cite{Bachler21}, an analysis by Shi and Tanaka using experimental measurements \cite{Shi20}, calculations based on molecular simulations with the two realistic empirical water models TIP4P/Ice and TIP4P/2005 \cite{Debenedetti20}, and a very recent extrapolation based on polynomial fits to equation of state data by Mishima and Sumita\cite{mishima2023equation}.
It follows from FIG.~\ref{fig:Fig3}B that, if such estimates are correct, all melting curves would be supercritical in experiments.
Furthermore, the relative positions of the ice polymorph melting curves and the critical point provided by SCAN in FIG.~\ref{fig:Fig3}A seems to be in excellent qualitative agreement with the experimental results shown in FIG.~\ref{fig:Fig3}B, i.e., the relative stability of all phases is captured qualitatively.
However, the quantitative positions of the melting curves and critical point in the $T-P$ plane differ significantly from experiments, which we attribute to the known limitations of SCAN \cite{Gartner20,Piaggi21}. We note that it is possible that SCAN somehow shifts the location of the critical point relative to the ice melting curves, however, given the qualitative correspondence between FIG.~\ref{fig:Fig3}A and FIG.~\ref{fig:Fig3}B, we do not expect this to be the case.
Moreover, the calculations described above based on a semi-empirical model also show that the melting line of ice V is supercritical, in disagreement with the original interpretation of the experiments and supporting the picture provided by the SCAN functional. 

In FIG.~\ref{fig:Fig3}B we have combined experimental melting curves for heavy water \cite{Mishima00} with estimates of the critical point based on experiments carried out using light water \cite{Bachler21,Shi20} and simulations that ignore nuclear quantum effects \cite{Debenedetti20}.
A figure equivalent to FIG.~\ref{fig:Fig3}B, replacing the melting curves of heavy water ice polymorphs with melting curves of light water 
 ices \cite{mishima2021liquid} is shown in the Supplementary Material \cite{SI}.
The isotopic effect in the melting lines is rather small, with melting temperatures of heavy water around 5 K higher than in light water \cite{mishima2021liquid}.
On the other hand, the isotopic effect on the location of the critical point has recently been estimated by Eltareb et al.~\cite{eltareb2022evidence} using path integral molecular dynamics and a semiempirical model of water.
They found a critical point location for heavy water 18 K and 9 MPa higher than in light water.
The combined isotopic effect on the melting curves and the location of the critical point may lead to a relative shift of around 12 K in light water compared to heavy water.
Therefore, isotopic effects are unlikely to affect the picture shown in FIG.~\ref{fig:Fig3}.
We also stress that our simulation results shown in FIG.~\ref{fig:Fig3}A ignore nuclear quantum effects.
They are thus more representative of heavy water than of light water.

The discrepancy between our simulation results and Mishima's experiments lead to the question of why a sharp discontinuity in slope was observed in the experimental melting curves for ice V and ice IV. Such behavior could perhaps be explained by immediate crystallization of ice I rather than melting to a metastable (relaxed) liquid state, which of course is not an issue in the simulations due to the separation of time scales of ice nucleation and liquid-like equilibration/relaxation. In this context, it should be noted that Mishima's hypothesized liquid-liquid phase transition is located very close to the homogeneous nucleation locus. Furthermore, the behavior reported by Mishima for the melting curves past the hypothesized LLT\cite{Mishima00} is remarkably noisy on the low-pressure side. Experimental studies explicitly targeted towards this issue are needed to definitively evaluate this hypothesis.

\section*{Conclusions}

Our results suggest that experiments reported by Mishima and Stanley that pointed to the existence of a liquid-liquid critical point at $\sim$0.1 GPa and $\sim$220 K \cite{Mishima98}, and subcritical melting curves for ice IV, V, and XIII \cite{Mishima00}, might call for a different interpretation.
While our first principles calculations do support the existence of a liquid-liquid critical point \cite{Gartner22}, they suggest its location to occur at lower temperatures than had been hitherto assumed, such that the melting curves of ice III, IV, V, VI, and XIII are in reality supercritical.
The relative stability of phases reported here is in excellent agreement with experiments, yet from a quantitative point of view our simulations are limited by the accuracy of our chosen semilocal DFT functional.
Future work could test our findings using more sophisticated DFT functionals or higher levels of electronic-structure theory.
Considering the plethora of known ice polymorphs, and the ones that continue to be discovered and characterized \cite{gasser2021structural}, the search for ices with subcritical melting curves may be a fruitful endeavor.
We also hope that our work will stimulate further experimental efforts to elucidate the behavior of melting curves in the vicinity of the liquid-liquid critical point and definitively explain the discrepancies between the experimental and computational results.

%

\section*{Methods}
\label{sec:methods2}

\subsection*{Molecular dynamics simulations}
We performed molecular dynamics simulations with the engine \textsc{LAMMPS} \citesupp{Plimpton95,thompson2022lammps} augmented by the \textsc{DeePMD-kit} \cite{Wang18,Zhang18}\citesupp{Zhang18end}.
In all simulations we used a time step for the integration of the equations of motion of 0.5 fs and the mass of hydrogen was set to 2 AMU so as to allow a longer integration step.
We maintained constant temperature using the stochastic velocity rescaling algorithm \citesupp{Bussi07} with a relaxation time of 0.1 ps.
We used a Parrinello-Rahman \citesupp{Parrinello81} type barostat with a relaxation time of 1 ps for pressure control.

\subsection*{Integration of the Clausius-Clapeyron equation}

We employed system sizes of 192, 324, 1024, 336, 640, and 336 molecules for bulk water, ice III, ice IV, ice V, ice VI, and ice XIII, respectively.
We obtained configurations with realistic proton disorder for all ice polymorphs using \textsc{GenIce} \citesupp{Matsumoto17}.
For the bulk liquid we used an isotropic barostat while for all ice polymorphs we employed a fully anisotropic barostat.
We integrated Eq.~\eqref{eq:Clausius-Clapeyron} numerically using a fourth-order Runge-Kutta algorithm.
The starting point for the integration and other computational details are described in the Supplementary Material \cite{SI}.
The relaxation time of liquid water increases dramatically for thermodynamic conditions below the Widom line, and we meticulously checked for adequate convergence of the average enthalpy and volume by performing relatively long simulations of up to 100 ns per state point.

\subsection*{Biased coexistence simulations}

We used system sizes of 648, 256, 672, and 672 water molecules for ice III, IV, V, and XIII, respectively.
The system size for ice IV is smaller than for other ices because the orthogonalized cell available in \textsc{GenIce} for ice IV contained 128 molecules.
For this reason, one must choose between a simulation box of 256 or 2048 molecules for ice IV coexistence simulations.
Due to the smaller system size used for ice IV, we also performed simulations at one pressure using a much larger system of 2048 water molecules.
The calculated finite size effect amounted to a 7 K increase in the melting temperature of the small system relative to that of the large system (see Supplementary Material \cite{SI}).
Therefore, the melting temperatures reported herein for the small ice IV system were correspondingly decreased by 7 K to take into account finite-size effects.
Based on the results of ref.~\cite{Piaggi21}, we expect finite-size effects in the melting curves of other polymorphs to be around 2 K, which is similar to the statistical uncertainty in the calculation. 
We ensured that both the liquid and ice were subject to the desired pressure by fixing the box dimensions parallel to the interface to the equilibrium value for the crystal and applying a barostat to the perpendicular direction.
We constructed the bias potential for the biased coexistence simulation using the On-the-fly Probability Enhanced Sampling (OPES) method \citesupp{Invernizzi20} as implemented in \textsc{PLUMED} \citesupp{Tribello14}.
The bias potential was a function of collective variables tailored to target each polymorph following Bore et al. \cite{Bore22}.
The collective variables represent the number of molecules with an ice-like environment and are based on a measure of similarity between environments in the target crystal structure and in the simulation box as described by Piaggi and Parrinello \citesupp{Piaggi19b}.
In FIG.~\ref{fig:Fig2}E we show the environments employed in the definition of the collective variable for ice III.
We used a uniform multiumbrella distribution \citesupp{Invernizzi20} for OPES with lower bound $N_l\:n$ and upper bound $N_l\:(n+1)$ where $N_l$ is the number of atoms in a layer and $n$ is an integer.
In this way, the growth and melting of a full layer of ice is sampled reversibly.
Further details of this methodology can be found in the Supplementary Material \cite{SI} and ref.~\cite{Bore22}.

\section*{Data availability}
All input files to reproduce the simulations are available for download at the Princeton DataSpace repository \url{https://doi.org/10.34770/pbja-we49}, and in the PLUMED NEST\citesupp{Bonomi19}, the public repository of the PLUMED consortium, as \href{https://www.plumed-nest.org/eggs/23/004/}{plumID:23.004}.

\section*{Code availability}
LAMMPS, Plumed, and the DeepMD-kit are free and open source codes available at \url{https://
lammps.sandia.gov}, \url{https://www.plumed.org}, and \url{http://www.deepmd.org}, respectively.

\section*{Acknowledgements}

We are grateful to Jack Weis for providing thermodynamic data of the liquid-liquid transition in the TIP4P/Ice model of water.
This work was conducted within the center Chemistry in Solution and at Interfaces funded by the USA Department of Energy under Award DE-SC0019394.
Simulations reported here were substantially performed using the Princeton Research Computing resources at Princeton University which is consortium of groups including the Princeton Institute for Computational Science and Engineering and the Princeton University Office of Information Technology’s
Research Computing department.

\section*{Author contributions}
P.G.D. conceived the project; P.M.P and T.E.G. performed research and wrote the original draft; P.M.P, T.E.G., R.C., and P.G.D. designed research, discussed results, and reviewed and edited the manuscript.

\end{document}